\documentstyle[12pt,aasms]{article}
\begin{document}
\title{Variations in Solar Luminosity from Time Scales of
Minutes to Months}
\author{Jon D. Pelletier\altaffilmark{1}}
\altaffiltext{1}{Department 
of Geological Sciences, Snee Hall, Cornell University,
Ithaca, NY 14853}
\begin{abstract}
We present the power spectrum of solar irradiance during 1985 and
1987 obtained from
the ACRIM project from time scales of minutes to months. At low frequency the
spectra are Lorentzian (proportional to $1/(f^{2}+f_{o}^{2})$).
At higher frequencies they are proportional to
$f^{-\frac{1}{2}}$. A linear, stochastic model of the turbulent heat transfer
between the granulation layer (modeled as a homogeneous thin layer with a
radiative boundary condition) and the rest of the convection zone (modeled
as a homogeneous thick layer with thermal and diffusion constants
appropriate  the lower convection zone)
predicts the observed spectrum.
\end{abstract}

\clearpage
\section{Introduction}
The luminosity of the sun has significant
variations on time scales from minutes to
years. Data from the NIMBUS 7 and ACRIM projects have provided us with a high
quality time series of these variations. Some aspects of these variations
can be attributed to specific physical processes
(see Stix 1989 for an introductory review). For example, 
at very high frequencies,
oscillations of the sun result in a well-understood five minute periodicity.
At the yearly and decadal
time scale there is a strong correlation
between the irradiance and variations in the solar magnetic
activity. Variations from
minutes to months have a simple spectral form but are not well understood
(Frohlich 1993).

Frohlich (1987) has published power spectra of ACRIM data from 1980 and
1985. He reported
that the power spectrum is flat for frequencies corresponding to time scales
greater than one week, proportional to $f^{-2}$ for time scales between
twelve hours and one week, and proportional to $f^{-1}$ at time scales down to
minutes. Our studies agree that the low frequency spectrum is Lorentzian.
We find, however, that the high-frequency spectrum is proportional to
$f^{-\frac{1}{2}}$ at time scales shorter than one day. 

Kuhn, Libbrecht,
\& Dicke (1988)
have suggested the possibility that the interaction between
the solar surface and deeper portions of the convection zone may cause the
low-frequency variations. We explore that possibility in this {\it Letter\/}.

If turbulent transfer in a system is dominated by eddies much smaller than
the system size, random convective action of turbulent eddies will be
analogous to the molecular agitation responsible for molecular diffusion
(Moffatt 1983).
In that approximation, turbulent transfer can be modeled
as a stochastic diffusion process. Random forcing of a deterministic
diffusion equation enables analytic study of the fluctuations from
equilibrium resulting from the stochasticity of turbulent transfer in any
geometry with any linear boundary condition. In this paper we study the
fluctuations in luminosity of a thin homogeneous surface granulation layer
with a linear radiation boundary condition exchanging heat with a
deep, homogeneous
layer below with density and diffusion constants appropriate to the lower
convection zone. The high and low crossover frequencies in the spectrum
correspond to time scales of thermal and radiative equilibration
of the convection
zone, respectively. The time scales for equilibration are given by the model
as a function of thermal and diffusion constants of the granulation layer
and lower
convection zone. Estimates of these constants with mixing length theory
yields order-of-magnitude agreement between the crossover frequencies
predicted by the model and those observed.

\section{Power Spectrum of ACRIM Data}
In Figure 1 we present the logarithm of the normalized
Lomb periodograms of ACRIM
solar irradiance data sampled during
1987 and 1985 plotted as a function of the logarithm of the frequency.
We chose to analyze these years since they appear to represent extremes
in the 
variation of solar activity at low frequencies. The low-frequency variance 
in the 1985 and 1987 data are small and large, respectively. 
Variations in the solar irradiance at yearly
time scales are generally agreed to be the result of variations
in magnetic activity.
We have chosen these years in 
order to assess the influence of magnetic activity on the power spectrum  
and distinguish its influence from that of the mechanism proposed in our 
model. 

Since the data were sampled at irregular intervals, simple FFT methods
of estimating the power spectrum are only available if we average the data
over some uniform time interval. We chose instead to use the Lomb
periodogram suggested by Press et al. (1992) for unevenly sampled data.
Above frequencies of $\log f=-2.0$ we averaged the
periodogram in logarithmically spaced
frequency intervals of $\log f=0.01$ to reduce the scatter. We subtracted 
$\log S(f)$ by 2.5 to plot it on the same graph as the 1987 data.  

The high-frequency behavior is the same for both spectra. A $f^{-2}$ region
flattens out to a $f^{-\frac{1}{2}}$ at frequencies greater than 
$f\approx\frac{1}{1\ day}$. Our observation of a
$f^{-\frac{1}{2}}$ scaling region at high frequencies 
disagrees with Frohlich's (1987)
conclusion that the high-frequency scaling region
is proportional to $f^{-1}$. He reported this conclusion for 1985 ACRIM
data, the same data we analyzed as part of our study.
Our confidence in our interpretation lies in the greater resolution of our
spectra, whose $f^{-\frac{1}{2}}$ range is $50 \%$ larger than Frohlich's
(1987).   
Large peaks
appear at the orbital frequency of the satellite and its harmonics. These
peaks are an artifact of the spectral estimation.

The low-frequency behavior of both spectra are Lorentzian (constant at low
frequencies and proportional to $f^{-2}$ at higher frequencies) in
agreement with Frohlich's (1987) results. Aside from the basic form of the 
spectra, there is a large variability between the crossover frequencies and
the magnitude of the two spectra reported here and with those published 
by Frohlich (1987). This variability was also discussed by Frohlich (1987).
The crossover frequency of the Lorentzian portion of the 1987 and 1985
data reported here are $f=\frac{1}{5\ months}$ and $f=\frac{1}{1\ month}$,
respectively. 
We interpret this variability as due to either variations in the magnetic
activity or to limitations of our model at these time scales. This 
{\it Letter\/} leaves this an open question. 

\section{Model of Variations in Solar Luminosity}
The
variations in the irradiance of the sun will be proportional to the
variations in its surface temperature. This follows from the fact that the
power emitted by the sun, modeled as a blackbody, 
$F-F_{e}=\sigma T^{4}- \sigma T_{e}^{4}$,
can be well approximated
by a linear dependence on $T-T_{e}$ for small departures from equilibrium.

Turbulent transport of heat in the convection zone of the sun can be modeled
by a stochastic diffusion process.
A stochastic diffusion process can be studied analytically by adding a noise
term to the flux of a deterministic diffusion equation (van Kampen 1981):
\begin{equation}
\rho c \frac{\partial \Delta T}{\partial t}=-\frac{\partial J}{\partial x}
\end{equation}
\begin{equation}
J=-\sigma\frac{\partial \Delta T}{\partial x} + \eta(x,t)
\end{equation}
where $\Delta T$ are the fluctuations in temperature from equilibrium and
the mean and variance of the noise is given by
\begin{equation}
\langle \eta(x,t)\rangle=0
\end{equation}
\begin{equation}
\langle \eta(x,t)\eta(x',t')\rangle\propto\sigma(x)\langle T(x)
\rangle^{2}\delta(x-x')\delta(t-t')
\end{equation}
Methods of studying transport by adding a noise term to a deterministic
differential equation are termed Langevin methods. Langevin methods have
been widely used to study the turbulent
diffusion of a passive scalar (temperature
or contaminant concentration) in the Earth's atmosphere and oceans
(see Csanady 1980 for a review). 

A diffusion process has a
frequency-dependent correlation length
$\lambda=(\frac{2D}{f})^{\frac{1}{2}}$ (Voss \& Clarke 1976).
At very high frequencies the
correlation length is smaller than the convection zone. In that scaling
region, the fluctuations in the temperature of the granular layer of
the sun have a power spectrum proportional to $f^{-\frac{1}{2}}$. To show
this, we present an argument due to Voss \& Clarke (1976).
The Fourier transform of the driven diffusion equation presented above is
\begin{equation}
\Delta T(k,\omega)=\frac{ik\eta (k,\omega)}{Dk^{2}-i\omega}
\end{equation}
The frequency-dependent correlation function,
$c_{T}(s,\omega)=<\Delta T(x+s,\omega) \Delta T^{*}(x,\omega)>$,
is then given by (Voss \& Clarke 1976)
\begin{equation}
c(s,\omega)\propto \frac{e^{-\frac{s}{\lambda}}}{\omega^{\frac{1}{2}}}
\cos (\frac{\pi}{4}+\frac{s}{\lambda})
\end{equation}
The power spectrum of fluctuations in the average temperature in a layer
of width $l$ will be a double spatial integral of the correlation function over
pairs of points in the volume of length $l$
and a time integral of the integrand multiplied
times a factor $\cos (\omega\tau)$ (the Wiener-Khintchine theorem). For low
frequencies, the factor $\cos (\omega\tau)$ is one so that the power spectrum
reduces to the spatial integral
\begin{equation}
S_{T}(\omega)\propto\int_{0}^{l}dx_{1}\int_{0}^{l}dx_{2}c_{T}(x_{1}-x_{2},
\omega)
\end{equation}
Since the correlation function is independent of $x_{1}-x_{2}$ for low
frequencies, integration will yield the same frequency dependence. Thus,
$S_{T}(\omega)\propto \omega^{-\frac{1}{2}}$.

At lower frequencies, the entire convection zone achieves thermal
equilibrium. The variance in temperature of the convection zone (and therefore
the irradiance) will now be determined by the radiation boundary condition.
The fluctuating heat transport (a consequence of the stochasticity of
turbulence) near the surface adds and subtracts heat from the top of the
convection zone. This results in temperature and irradiance variations
with a random walk
($f^{-2}$) spectrum.

The fluctuating input and output of heat in the $f^{-2}$ region will
cause large variations from equilibrium. When the temperature of the
convection zone becomes larger than the equilibrium temperature, it will
radiate, on average, more heat than at equilibrium. Conversely, when the
temperature of the convection zone wanders lower than the equilibrium
temperature, less heat is radiated. This negative feedback
limits the variance at low frequencies resulting in a constant power spectrum.

The Lorentzian portion of the spectrum has exactly the same physics as
shot noise in an RC circuit. In shot noise, the Brownian motion of electrons
gives rise to a $f^{-2}$ spectrum of charge stored on the capacitor. The
frequency-dependent reactance of the capacitor limits the variance of the
charge through a transient response $\frac{dQ}{dt}=\frac{Q}{\tau_{o}}$
where $\tau_{o}=2\pi RC$. This response is equivalent to the linear radiation
boundary condition we have imposed at the surface of the sun. In our model,
two homogeneous layers with different thermal and diffusion constants
comprise the interacting ``circuit'' elements. The crossover frequency is
given by $f_{o}=\frac{1}{2\pi R_{eq}C_{eq}}$ where $R_{eq}$ and $C_{eq}$
are the equivalent thermal resistance (inverse of the thermal conductivity) and
the thermal capacitance of the parallel combination of the two layers,
respectively.

The model we present was solved in the context of a different problem by
van Vliet, van der Ziel, \& Schmidt (1980). They considered the temperature
fluctuations in a thin metal film supported by a substrate.

The geometry of the model is a thin granulation layer of width
$2$x$10^6$ m and uniform density
(equal to the density at the bottom of the granulation layer where most
of the heat capacity resides) of $0.003$ kg/m$^{3}$ coupled to a thick layer
of uniform density representing the rest of the convection zone. The layers
have a planar geometry for convenience in this simplified model.
The turbulent diffusivity
is estimated from mixing length theory as $\alpha=\frac{1}{3}vl$ where $v$ and
$l$ are the characteristic velocity and eddies sizes, respectively. The
eddy size, $l$, is usually approximated as one pressure scale height. In the
case of the granulation layer, however, the dominant eddy size is the size
of the convection cell, approximately $2$x$10^{6}$ m.
The velocity at the bottom of
the granulation layer is on the order of $1000$ m/s. These estimates
yield an eddy diffusivity of $10^{9}$ m$^{2}$/s near the solar surface.
The thermal conductivity,
$\sigma =\alpha\rho c$, is $3$x$10^{7}$ W/m$^{o}$K since $c=10$ J/kg$^{o}$K
for a monatomic hydrogen gas.
The velocity, thickness, and specific heat values are from a standard
solar model presented in Stix (1992).
The density values are from Bohm (1963).

The granulation layer sits atop the rest of the convection zone.
We approximate the density, width, and diffusivity of the remainder of the
convection zone by their values near the bottom of the convection zone
because
its high density results in a concentration of heat capacity there and
its slow diffusivity is the rate-limiting step of thermal
equilibration
of the convection zone. The width and dominant eddy scale will both be
given by $10^{7}$ m, the width of the lowest pressure scale height of the
convection zone. The density is estimated to be $0.5$ kg/m$^{3}$ and
the velocity is on the order of $20$ m/s. These values are the
arithmetic means of the densities and velocities at the top and bottom
of the pressure scale height. These values yield an eddy diffusivity of
$10^{8}$ m$^{2}$/s for the bottom of the convection zone. Our diffusivities
agree with the estimates of Stix (1992) who quote the range of
diffusivities in the convection zone as $10^{8}-10^{9}$ m$^{2}$s$^{-1}$.
The thermal conductivity is $3$x$10^{8}$ W/m$^{o}$K.

The equation for temperature fluctuations in space and time in the model
is
\begin{equation}
\frac{\partial \Delta T(x,t)}{\partial t}-\alpha (x)
\frac{\partial ^{2} \Delta T(x,t)}
{\partial x^{2}}
=-\frac{\partial\eta(x,t)}{\partial x}
\end{equation}
with
\begin{equation}
\langle \eta(x,t)\rangle=0
\end{equation}
\begin{equation}
\langle \eta(x,t)\eta(x',t')\rangle\propto\sigma(x)\langle T(x)
\rangle^{2}\delta(x-x')\delta(t-t')
\end{equation}

The boundary conditions are that there be no heat flow out of
bottom of the convection zone and continuity of temperature
and heat flux at the boundary separating the granulation layer and the
deeper convection zone:
\begin{equation}
\sigma '
\frac{\partial T}{\partial x}|_{x=w_{2}} = 0
\end{equation}
\begin{equation}
\Delta T(x=w_{1}^{+})=\Delta T(x=w_{1}^{-})
\end{equation}
\begin{equation}
\sigma \frac{\partial \Delta T}{\partial x}|_{x=w_{1}^{-}}=
\sigma ' \frac{\partial \Delta T}{\partial x}|_{x=w_{1}^{+}}
\end{equation}
where $w_{1}$ and $w_{2}$ are the widths of the granulation layer and
deep convection zone, respectively and the primes denote thermal and
diffusion constants of the deep convection zone.

At the top of the granulation layer
we impose a blackbody radiation boundary
condition.
The heat emitted from the granulation layer is dependent on the
temperature of the layer through the Stefan-Boltzmann law. Since
solar surface temperature variations are small, within a linear
approximation, the emitted temperature will be
proportional to the temperature difference from equilibrium
(the same approximation is commonly made in analytic models of
climate change, i.e. Ghil 1983).
At the solar surface, then,
\begin{equation}
\sigma \frac{\partial \Delta T}{\partial x}|_{x=0}=g\Delta T(x=0)
\end{equation}
where $g=4\sigma _{b}T_{o}^{3}=2$x$10^{4}$ W/m$^{2}$$^{o}$K
is the thermal conductance of heat out of the sun and $\sigma _{b}$ is the
Stefan-Boltzmann constant.

van Vliet et al. (1980) used Green's functions to solve this model.
The Green's function of the Laplace-transformed
diffusion equation is defined by
\begin{equation}
i\omega G(x,x',i\omega)-\alpha (x)\frac{\partial^{2}G(x,x',i\omega)}
{\partial x^{2}}=\delta (x-x')
\end{equation}
where G is governed by the same boundary conditions as $\Delta T$.
This equation can be solved by separating $G$ into two parts: $G_{a}$ and
$G_{b}$ with $x < x'$ and $x > x'$, respectively, where $G_{a}$ and
$G_{b}$ satisfy the homogeneous (no forcing) diffusion equation with a
jump condition relating $G_{a}$ and $G_{b}$:
\begin{equation}
\frac{\partial G_{a}}{\partial x}|_{x=x'}-\frac{\partial G_{b}}{\partial x}
|_{x=x'}=\frac{1}{\alpha (x')}
\end{equation}

The power spectrum
of the average temperature in the granualtion layer
in terms of G is given by van Vliet et al. (1980) as:
\begin{equation}
S_{\Delta T_{av}}(f)\propto
Re (\int_{0}^{w_{1}}\int_{0}^{w_{1}} G_{1}(x,x',i\omega)
dx dx')
\end{equation}
\begin{equation}
\propto
Re(\int_{0}^{w_{1}}\int_{0}^{x}G_{1b}(x,x',i\omega)dx dx'+
\int_{0}^{w_{1}}\int_{x}^{w_{1}}G_{1a}(x,x',i\omega)dx dx')
\end{equation}
where $G_{1}$ stands for the solution
to the differential equation for G where the source point is located in
the granulation layer. Two forms of $G_{1a}$ and $G_{1b}$ are necessary for $x$
located above and below $x'$, respectively, due to the discontinuity
in the derivative of $G_{1}$ created by the delta function
(the jump condition).
The solution of $G_{1}$ which satisfies the above differential equation and
boundary conditions is
\begin{equation}
G_{1a}=\frac{L}{\alpha K}(\frac{\sigma 'L}{\sigma L'}\sinh(\frac{w_{1}-x'}{L})
\sinh(\frac{w_{2}}{L'})+\cosh(\frac{w_{1}-x'}{L})\cosh(\frac{w_{2}}{L'}))
(\sinh(\frac{x}{L})+\frac{\sigma}{Lg}\cosh(\frac{x}{L}))
\end{equation}
and
\begin{equation}
G_{1b}=G_{1a}+\frac{L}{\alpha}\sinh(\frac{x'-x}{L})
\end{equation}
where
\begin{equation}
K=(\sinh(\frac{w_{1}}{L})+\frac{\sigma}{Lg}\cosh(\frac{w_{1}}{L}))
\frac{\sigma 'L}{\sigma L'}\sinh(\frac{w_{2}}{L'})+(\cosh(\frac{w_{1}}{L})
+\frac{\sigma}{Lg}\sinh(\frac{w_{1}}{L}))\cosh(\frac{w_{2}}{L'})
\end{equation}
and $L=(\frac{\alpha}{i\omega})^{\frac{1}{2}}$ and
$L'=(\frac{\alpha '}{i\omega})^{\frac{1}{2}}$. Performing the integration
van Vliet et al. obtained
\begin{eqnarray}
S_{\Delta T_{av}}(f)\propto
Re (L^{2}(\frac{\sigma 'L}{\sigma L'}
\tanh(\frac{w_{2}}{L})((\frac{gw_{1}}{\sigma}-1)\tanh(\frac{w_{1}}{L})
-\frac{2gL}{\sigma}\frac{\cosh(w_{1}/L)-1}{\cosh(w_{1}/L)}+\frac{w_{2}}{L})
\nonumber\\
+(\frac{gw_{1}}
{\sigma}+(\frac{w_{1}}{L}-\frac{gL}{\sigma}\tanh(\frac{w_{1}}{L}))
((\tanh(\frac{w_{1}}{L})+\frac{\sigma L}{g})\frac{\sigma 'L}{\sigma L'}
\tanh(\frac{w_{2}}{L'})+(1+\frac{\sigma}{Lg}
\tanh(\frac{w_{1}}{L})))^{-1}
\label{bigone}
\end{eqnarray}
For very low frequencies,
\begin{equation}
\tanh(\frac{w_{1}}{L})\approx\frac{w_{1}}{L},
\tanh(\frac{w_{2}}{L'})\approx\frac{w_{2}}{L'}
\end{equation}
\begin{equation}
\frac{\cosh(w_{1}/L)-1}{\cosh(w_{1}/L)})\approx\frac{1}{2}\frac{w_{1}^{2}}{L^{2}
}
\end{equation}
Reducing eq. (\ref{bigone}),
\begin{equation}
S_{\Delta T_{av}}(f)\propto \frac{1}{1+\frac{\omega^{2}}{\omega_{0}^{2}}}
\propto \frac{1}{f^{2}+f_{0}^{2}}
\end{equation}
which is the low-frequency Lorentzian spectrum observed in the ACRIM data.
The crossover frequency as a function of the constants chosen
for the model is
\begin{equation}
f_{0}=\frac{g}{2\pi(w_{1}c\rho + w_{2}c'\rho '(1+\frac{g w_{1}}{\sigma}))}
\approx\frac{\sigma}{2\pi w_{1}w_{2}c'\rho '}\approx\frac{1}{8\ months}
\end{equation}
which is within an order of magnitude of the observed crossover frequencies
of the 1987 and 1985 ACRIM data, $f=\frac{1}{5\ months}$ and 
$f=\frac{1}{1\ month}$ respectively.

At low frequencies
\begin{equation}
\tanh(\frac{w_{1}}{L})\approx\frac{w_{1}}{L},
\tanh(\frac{w_{2}}{L'})\approx 1
\end{equation}
\begin{equation}
\frac{\cosh(w_{1}/L)-1}{cosh(w_{1}/L)}\approx\frac{1}{2}\frac{w_{1}^{2}}{L^{2}}
\end{equation}
then
\begin{equation}
S_{T_{av}}(f)\propto \frac{1}{2}(\frac{2gw_{1}}{\sigma})^{\frac{1}{2}}
(\frac{c\rho\sigma}{c'\rho '\sigma '})^{\frac{1}{2}}
(\frac{g}{w_{1}\rho c f})^{\frac{1}{2}}\propto f^{-\frac{1}{2}}
\end{equation}
as observed.

The high and low-frequency spectra meet at
\begin{equation}
f_{1}=\frac{g}{w_{1}\rho c}(\frac{\sigma}{2gw_{1}})^{\frac{1}{3}}
(\frac{c'\rho'\sigma'}{c\rho\sigma})^{\frac{1}{3}}
4^{\frac{1}{3}}(\frac{c\rho w_{1}}
{c'\rho ' w_{2}})^{\frac{4}{3}}
\end{equation}
\begin{equation}
\approx\frac{1}{6\ hours}
\end{equation}
which also agrees to within an order of magnitude with the crossover
frequency of the ACRIM data ($f=\frac{1}{1\ day}$).
 
We have applied the same model to climate change (Pelletier 1995).
The above model gives
the fluctuations in the average temperature in a thin, homogeneous layer
(Earth's atmosphere) coupled with another
homogeneous layer with different thermal
and diffusion constants (the ocean). As part of that work, we analyzed the
Vostok ice core. We found the same
spectral form reported here in the ACRIM data. The thermal and radiative
time scales in that dataset were $2,000$ and $40,000$ years, respectively.
Estimates of those time scales based upon thermal constants and diffusion
constants inferred from tracer studies in the atmosphere and ocean
matched the time scales of the crossover frequencies of the Vostok data well.

\section{Conclusions}
We have presented evidence that the power spectrum of variations in solar
irradiance exhibits four scaling regions. We presented a model
originally due to van Vliet et al. (1980)
proposed to study temperature fluctuations in a metallic film
(granulation layer) supported by a substrate (deep convection zone) 
that matches the observed
frequency dependence of the power spectrum of irradiance fluctuations.

\acknowledgements
I wish to thank Donald Turcotte and Ed Salpeter for helpful conversations
of related work. I am indebted to Sandy Kwan of JPL who provided me with
the ACRIM data.

\clearpage

\begin{figure}
\caption{
Figure 1: Logarithm of the normalized Lomb periodograms of solar
irradiance in 1987 (upper plot) and 1985 (lower plot) 
from the ACRIM project versus the logarithm of the
frequency in hours$^{-1}$. The crossover frequencies for 1987 are
$f_{0}=\frac{1}{5\ months}$ and $f_{1}=\frac{1}{1\ day}$. The crossover 
frequencies for 1985 are $f_{0}=\frac{1}{1\ month}$ and 
$f_{1}=\frac{1}{1\ day}$. The 1985 spectrum is shifted down by 
$\log S(f)=-2.5$.
}
\end{figure}
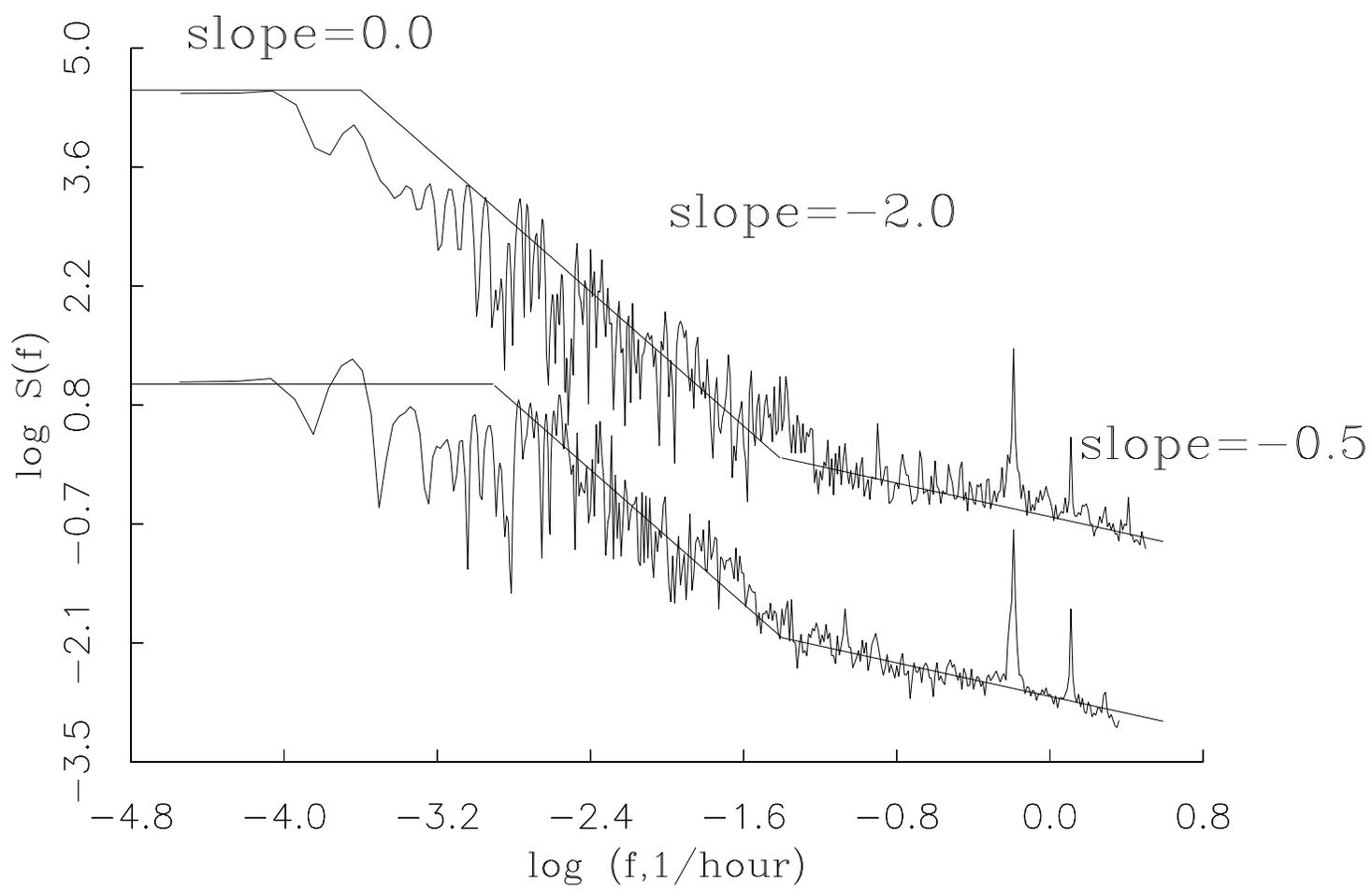
\end{document}